\documentclass[slac_one]{revtex4}
\usepackage{graphicx}
\usepackage{fancyhdr}

\renewcommand{\baselinestretch}{1.2}
\begin{document}
\begin{titlepage}

\thispagestyle{empty}
\def\thefootnote{\fnsymbol{footnote}}       

\begin{center}
\mbox{ }

\end{center}
\begin{flushright}
\Large
\mbox{\hspace{10.2cm} physics/0507028} \\
\end{flushright}
\begin{center}
\vskip 1.0cm
{\Huge\bf
Charge Transfer Inefficiency Studies 
}
\vspace{2mm}

{\Huge\bf
for CCD Vertex Detectors at a LC
}
\vskip 1cm
{\LARGE\bf Andr\'e Sopczak on behalf of the LCFI Collaboration}\\
\smallskip
\Large Lancaster University

\vskip 2.5cm
\centerline{\Large \bf Abstract}
\end{center}

\vskip 4.0cm
\hspace*{-0.5cm}
\begin{picture}(0.001,0.001)(0,0)
\put(,0){
\begin{minipage}{\textwidth}
\Large
\renewcommand{\baselinestretch} {1.2}
The Linear Collider Flavour Identification (LCFI) collaboration studies 
CCD detectors for quark flavour 
identification in the framework of a future linear e$^+$e$^-$ collider.
The flavour identification is based on precision reconstruction of charged 
tracks very close to the 
interaction point. Therefore, this detector will be exposed to a high level 
of radiation and thus an 
important aspect of the vertex detector development are radiation hardness 
studies. Results of detailed 
simulations of the charged transport properties of a CCD prototype chip are 
reported and compared with 
initial measurements. The simulation program allows to study the effect of 
radiation damage after the 
exposure of the detector to a realistic radiation dose, which is expected in 
the environment of detector operation at a future LC.
\renewcommand{\baselinestretch} {1.}

\normalsize
\vspace{4.5cm}
\begin{center}
{\sl \large
Presented at the 2005 International Linear Collider Workshop - Stanford, 
U.S.A., \\
to be published in the proceedings.
\vspace{-6cm}
}
\end{center}
\end{minipage}
}
\end{picture}
\vfill

\end{titlepage}

\newpage
\thispagestyle{empty}
\mbox{ }
\newpage
\setcounter{page}{0}

\newcommand{\cm}[1]{\ensuremath{ {{\rm cm}^{#1}}}}
\newcommand{\ntp}[2]{\ensuremath{#1\times10^{#2} } }
\newcommand{\asb}[2]{\ensuremath{#1_{\rm #2} }}

\title{{\small{2005 International Linear Collider Workshop - Stanford,
U.S.A.}}\\ 
\vspace{12pt}
Charge Transfer Inefficiency Studies for CCD Vertex Detectors at a LC} 

\author{Andr\'e Sopczak on behalf of the LCFI Collaboration}
\affiliation{Lancaster University, Lancaster, LA1 4YW, UK}

\begin{abstract}
The Linear Collider Flavour Identification (LCFI) collaboration studies CCD detectors for quark flavour 
identification in the framework of a future linear e$^+$e$^-$ collider.
The flavour identification is based on precision reconstruction of charged tracks very close to the 
interaction point. Therefore, this detector will be exposed to a high level of radiation and thus an 
important aspect of the vertex detector development are radiation hardness studies. Results of detailed 
simulations of the charged transport properties of a CCD prototype chip are reported and compared with 
initial measurements. The simulation program allows to study the effect of radiation damage after the 
exposure of the detector to a realistic radiation dose, which is expected in the environment of detector 
operation at a future LC.
\end{abstract}

\maketitle

\thispagestyle{fancy}

\section{INTRODUCTION} 
An important requirement of a vertex detector is to remain tolerant to radiation damage for its anticipated lifetime.  
Two different CCDs have been considered in this study.
The majority of simulations so far have been performed for the 3-phase CCD, CCD58, with  serial readout.
First data measurements have been performed on an unirradiated 2-phase column parallel CCD, CPC-1.

CCDs suffer from both surface and bulk radiation damage, however, when considering charge transfer losses in buried 
channel devices only bulk traps are important.
These defects create energy levels between the conduction and 
valance band, hence electrons may be captured by these new levels.
Captured carriers are also emitted back to the conduction band, but on a different time scale. 
For a signal packet this may lead to a decrease in charge as it is transfered to the output 
and may be quantified by its Charge Transfer Inefficiency (CTI), where a charge of amplitude $Q_0$ 
transported across $m$ pixels will have a reduced charge given by
\begin{eqnarray}
Q_{\rm m}=Q_0(1-{\rm CTI})^m.
\label{eqn:cti}
\end{eqnarray}
The CTI value depends on many parameters, some related to the trap characteristics such as:
trap energy level, capture cross-section, and trap concentration. Operating conditions also
affect the CTI as there is a strong temperature dependence on the trap capture rate and also a
variation of the CTI with the readout frequency. Other factors are also relevant, for example the
occupancy ratio of pixels, which influences the fraction of filled traps in the CCD transport region.
Dark current effects have been measured.

\section{SIMULATION}
\begin{figure}[htp]
\begin{minipage}[0.45]{5cm}
\includegraphics[height=6cm]{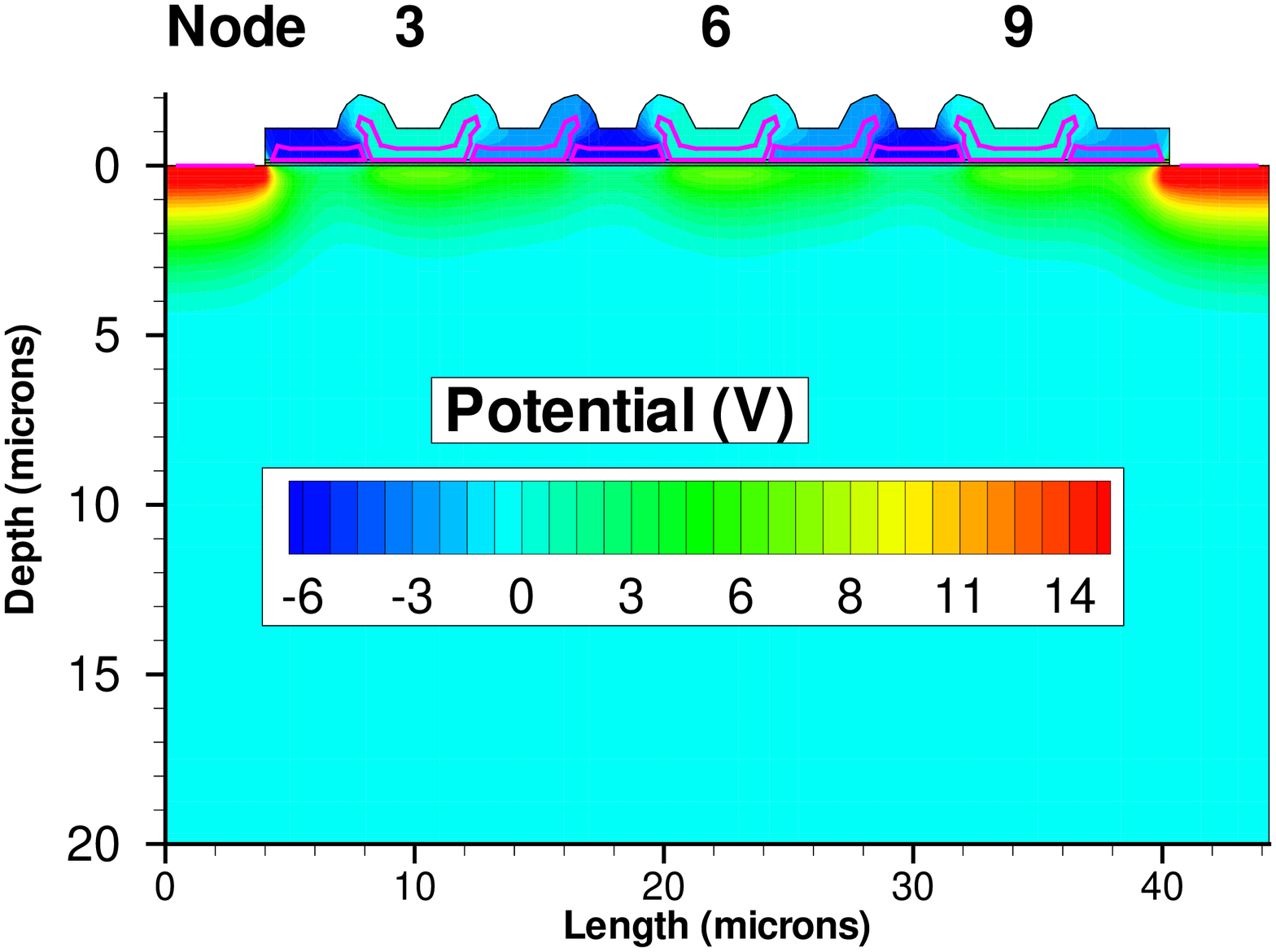}
\end{minipage} \hfill
\begin{minipage}[0.45]{8cm}
\begin{tabular}{l | l| l| l}
$\asb{E}{c} - \asb{E}{t}$ (eV) 
 & Type & $C$ (\cm{-3}) &$\sigma_{\rm n}$ (\cm{-2})\\ \hline
0.17 & Acceptor & \ntp{1}{11} & \ntp{1}{-14}\\
0.44 & Acceptor & \ntp{1}{11} & \ntp{1}{-15}
\end{tabular}
\vspace*{1cm}
\caption{\label{fig:trap}
Left: Detector structure and potential at gates (nodes) after initialization. The signal charge 
is injected under node 3.
Right: 
Energy levels $E$, trap concentrations $C$, and electron-capture cross-section $\sigma_{\rm n}$ 
used in simulation. }
\end{minipage}
\end{figure}

\begin{figure}[h]
\vspace*{-0.5cm}
\begin{minipage}[0.45]{8.8cm}
\includegraphics[width=9cm,height=6cm]{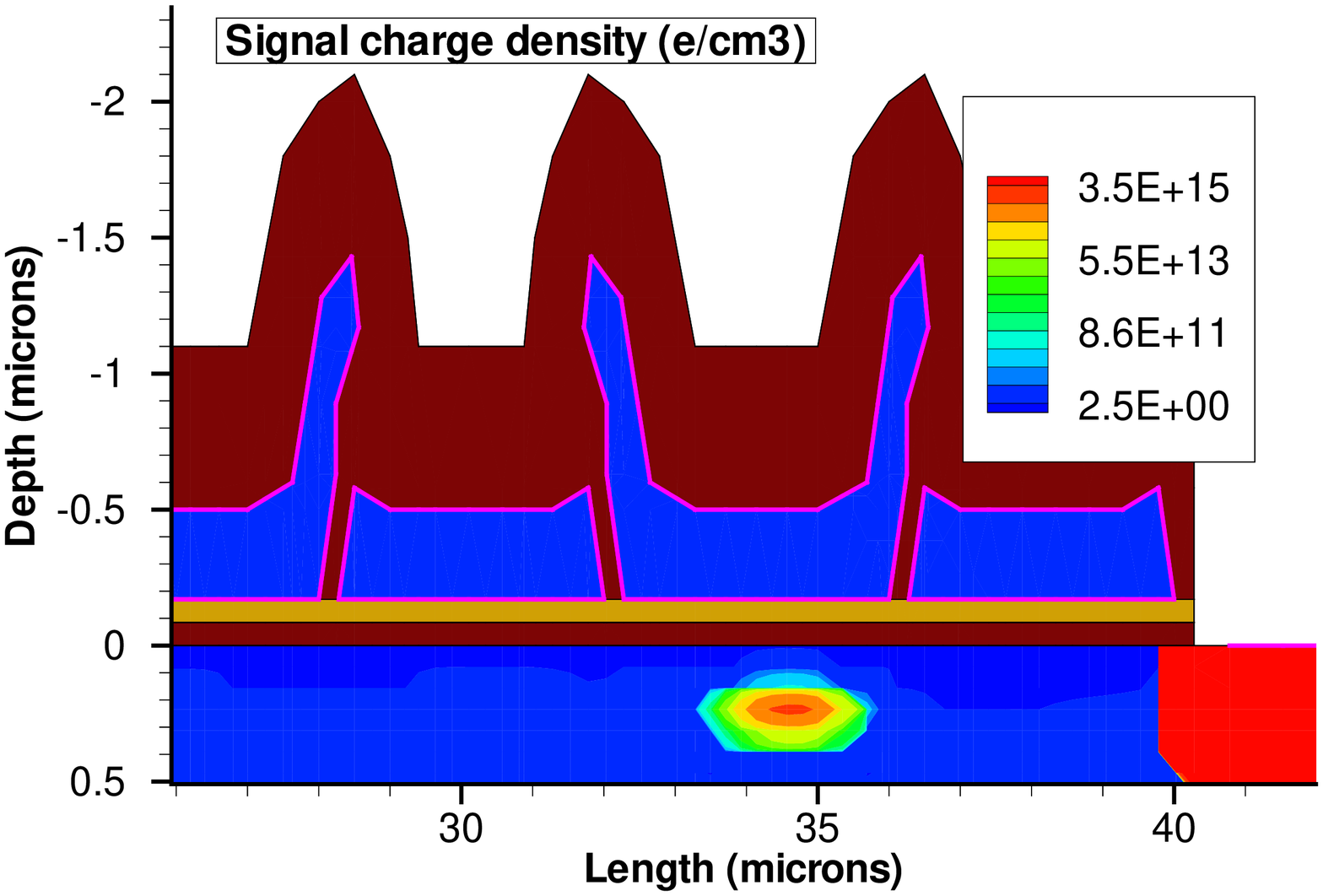}
\end{minipage} \hfill
\begin{minipage}[0.45]{8.8cm}
\includegraphics[width=9cm,height=6cm]{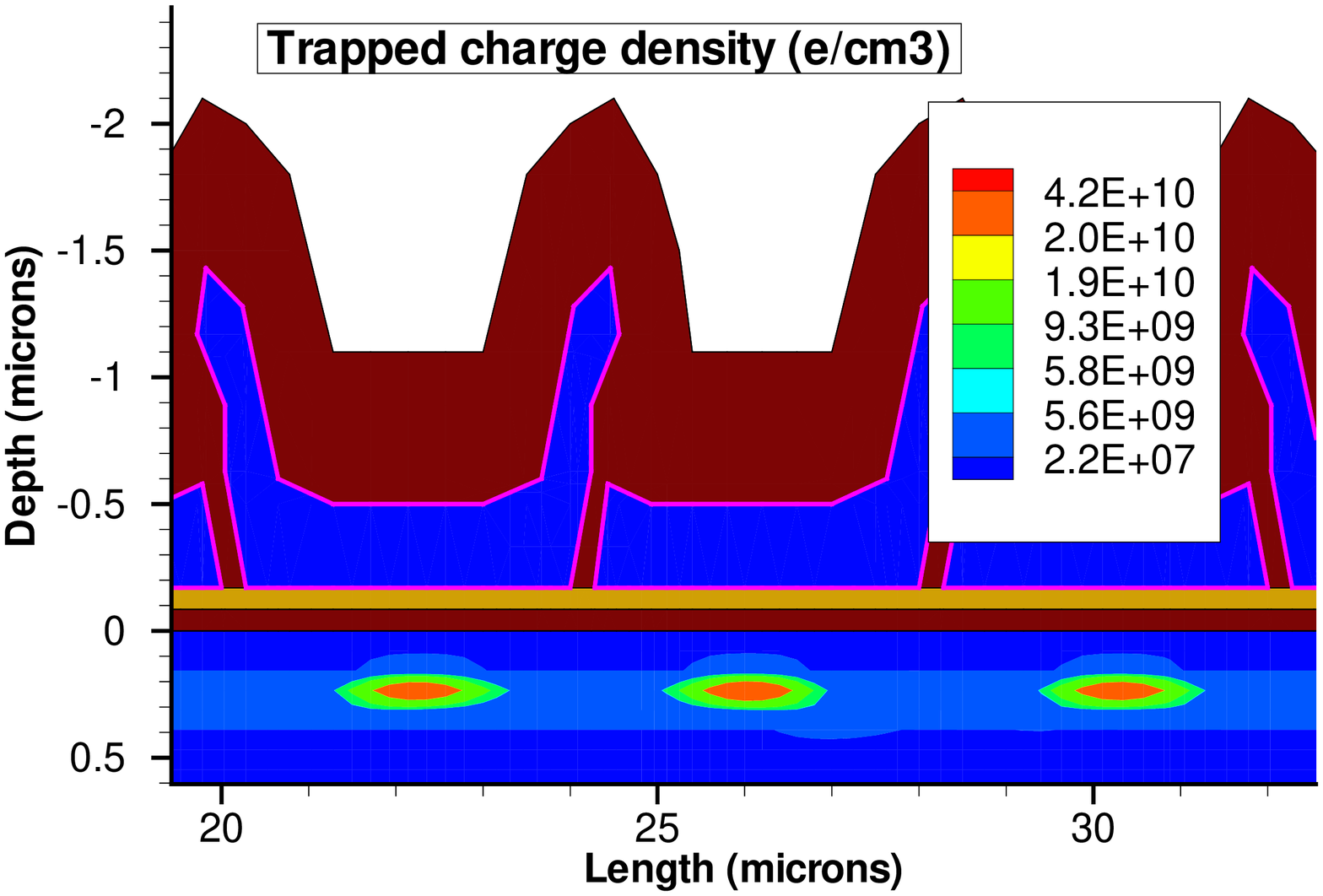}
\end{minipage}
\label{fig:transport}
\caption{Left: Signal charge density, almost at output gate.
Right: Trapped charge density, from transfer of signal charge. 
}
\end{figure}

The simulations with ISE-TCAD (version 7.5) are performed using a 2-dimensional model for a 3-phase CCD 
(Fig.~\ref{fig:trap}).
Parameters of interest are the readout frequency, up to 50 MHz, and the operating temperature 
between 100 K and 250 K. The charge in transfer and the trapped charge is shown in 
Fig.~\ref{fig:transport}. From the two traps considered only the 0.17 eV trap produced a non-negligible 
CTI for the explored parameter ranges. The signal charge used in the simulation represents a charge 
deposited from an Fe$^{55}$ source. The X-ray emission (mainly 5.9 keV) generates about 1620 electrons in the CCD, which is similar to the charge generated by a MIP.
The linearity of the CTI value with respect to the trap concentration was verified 
in the simulation.

\subsection{0.44 eV Trap CTI Contribution}
\begin{figure}[h]
\includegraphics[height=6.5cm,width=8.0cm]{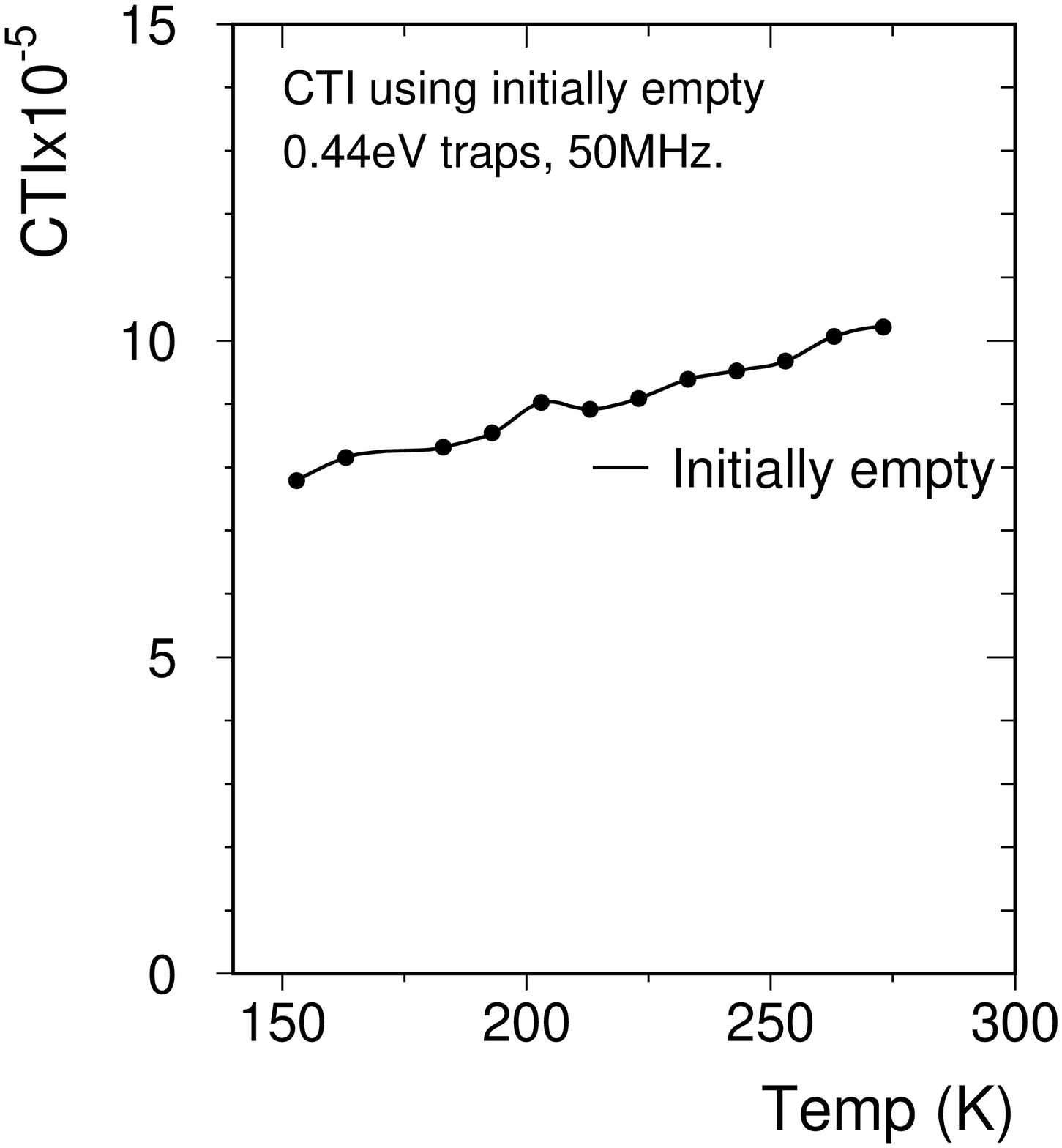} \hfill
\includegraphics[height=6.5cm,width=8.0cm]{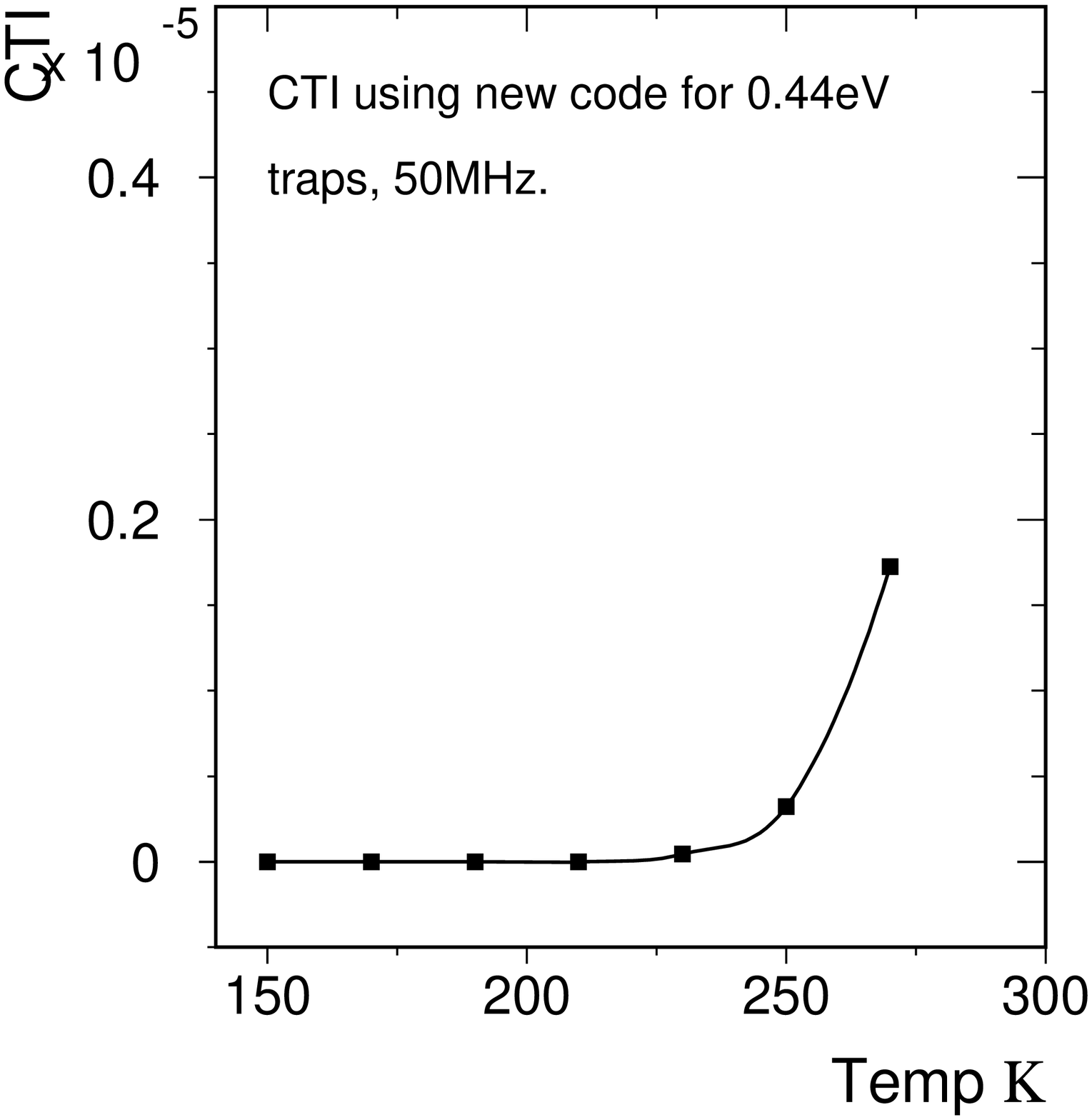}
\vspace*{-0.5cm}
\caption{\label{fig:trap044}
Left: CTI value for initially empty 0.44 eV traps.
Right: Partially filled 0.44 eV traps.
}
\end{figure}

Empty trap simulation may not be a good approximation. We consider partially filled traps
to improve the simulation by representing a continuous readout process. 
The results from the initially empty and partially filled traps are compared in Fig.~\ref{fig:trap044}.
A negligible contribution to the CTI from 0.44 eV trapping for partially filled traps 
(due to long emission time) are obtained. Thus, the 0.44 eV traps are ignored in further studies. 

\subsection{0.17 eV Trap CTI Contribution}
\begin{figure}[h]
\includegraphics[height=6.5cm,width=8.0cm]{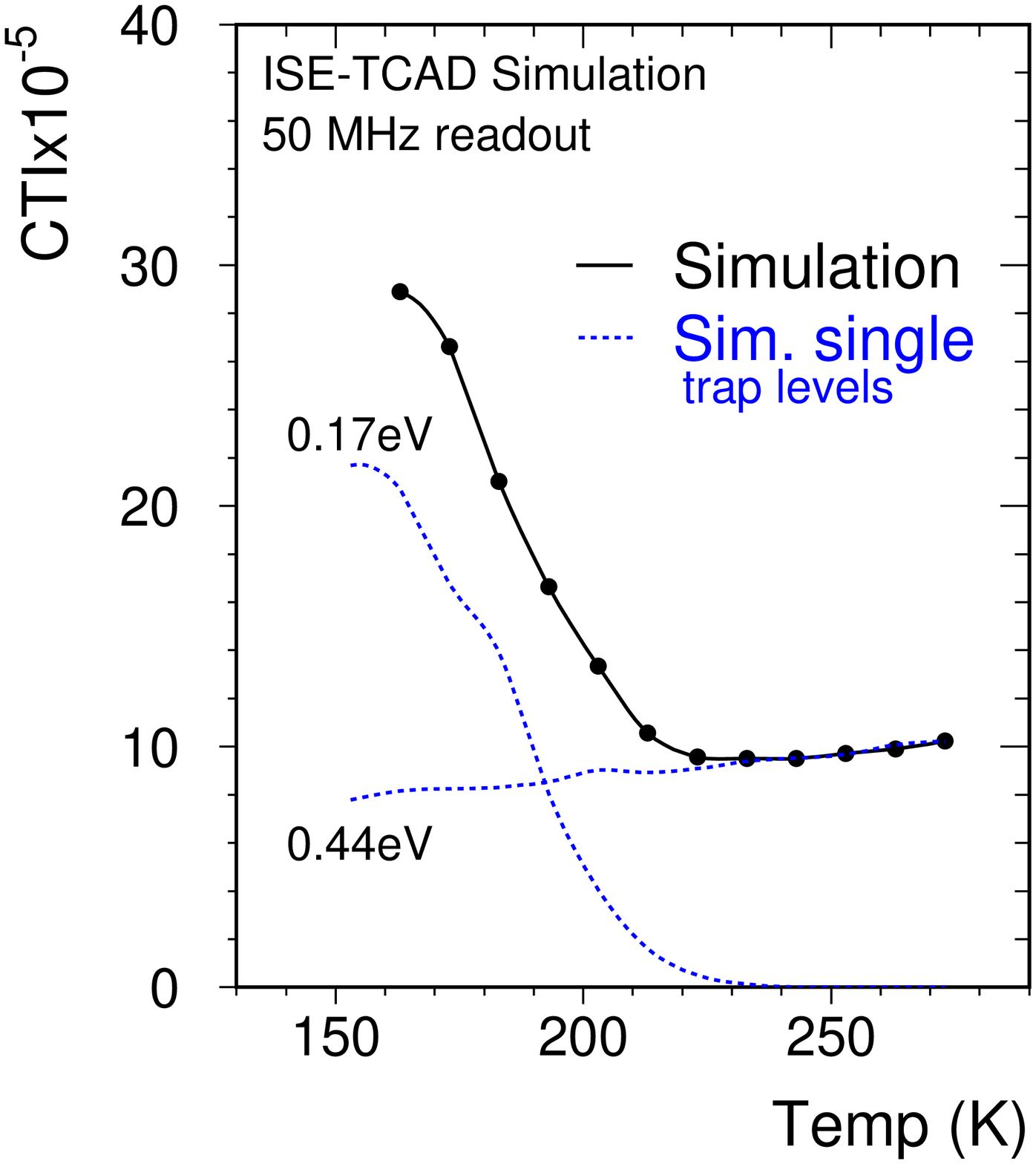} \hfill
\includegraphics[height=6.5cm,width=8.0cm]{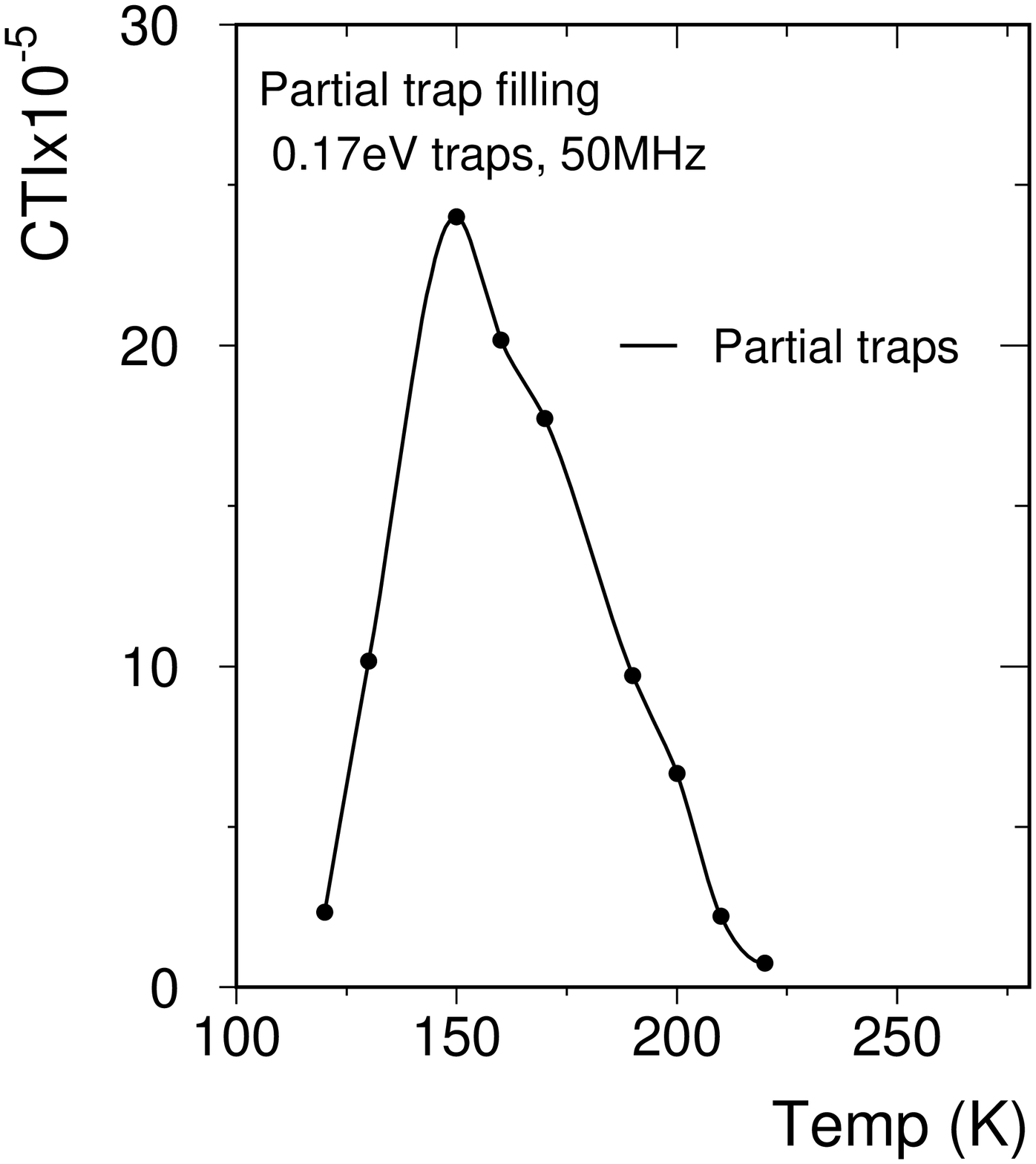}
\vspace*{-0.5cm}
\caption{\label{fig:trap017}
Left: CTI value for initially empty 0.17 eV traps, 0.44 eV traps and their sum.
Right: Partially filled 0.17 eV traps.
}
\end{figure}

Figure~\ref{fig:trap017} shows the CTI simulation for initially empty and partially filled traps. 
A clear peak structure is observed. New experimental data will cover the simulated temperature range.

\subsection{Frequency Dependence}
\begin{figure}[h]
\begin{minipage}[0.45]{8cm}
\includegraphics[height=6.5cm,width=8cm]{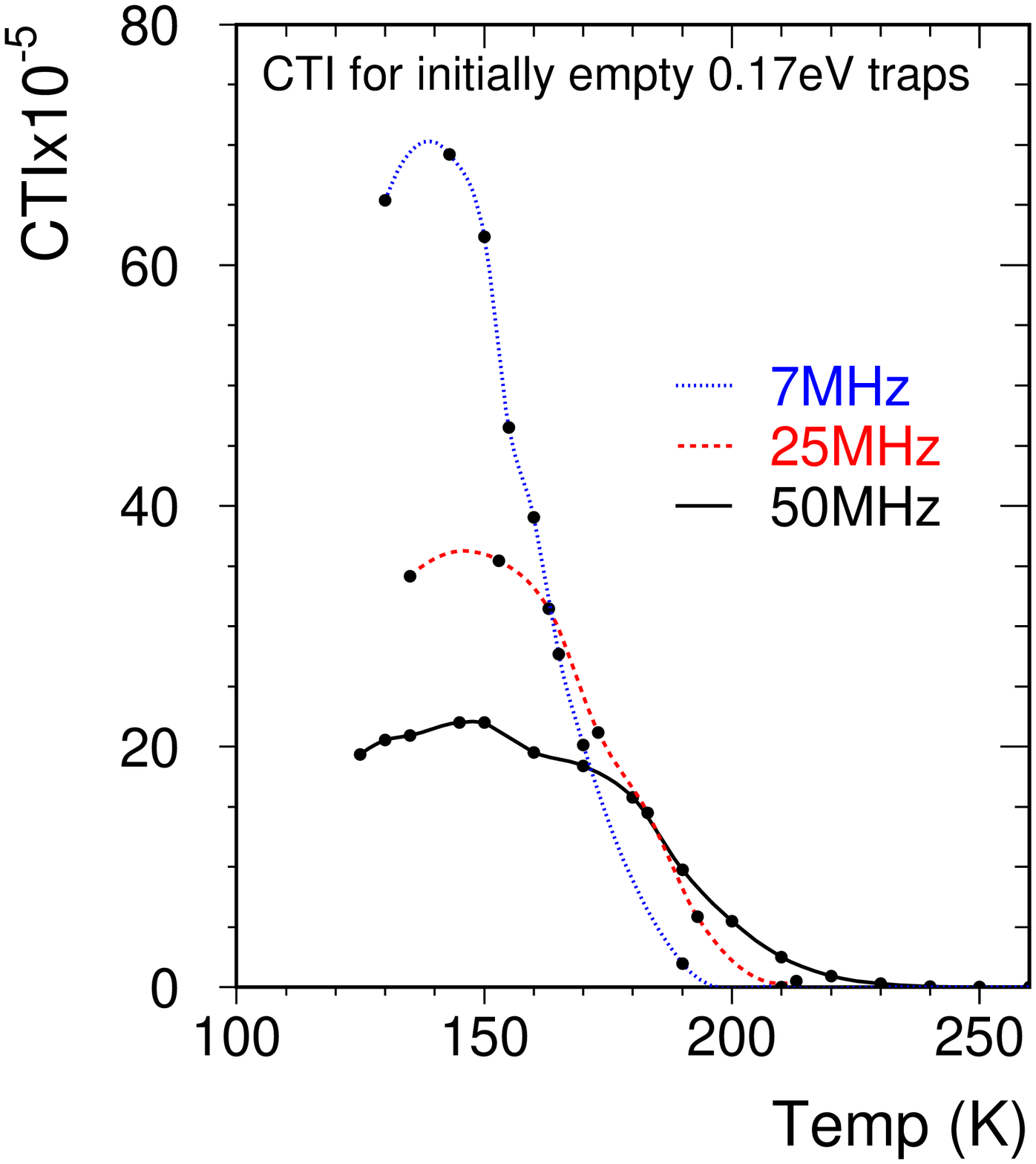}
\end{minipage} \hfill
\begin{minipage}[0.45]{8cm}
\includegraphics[height=6.5cm,width=8cm]{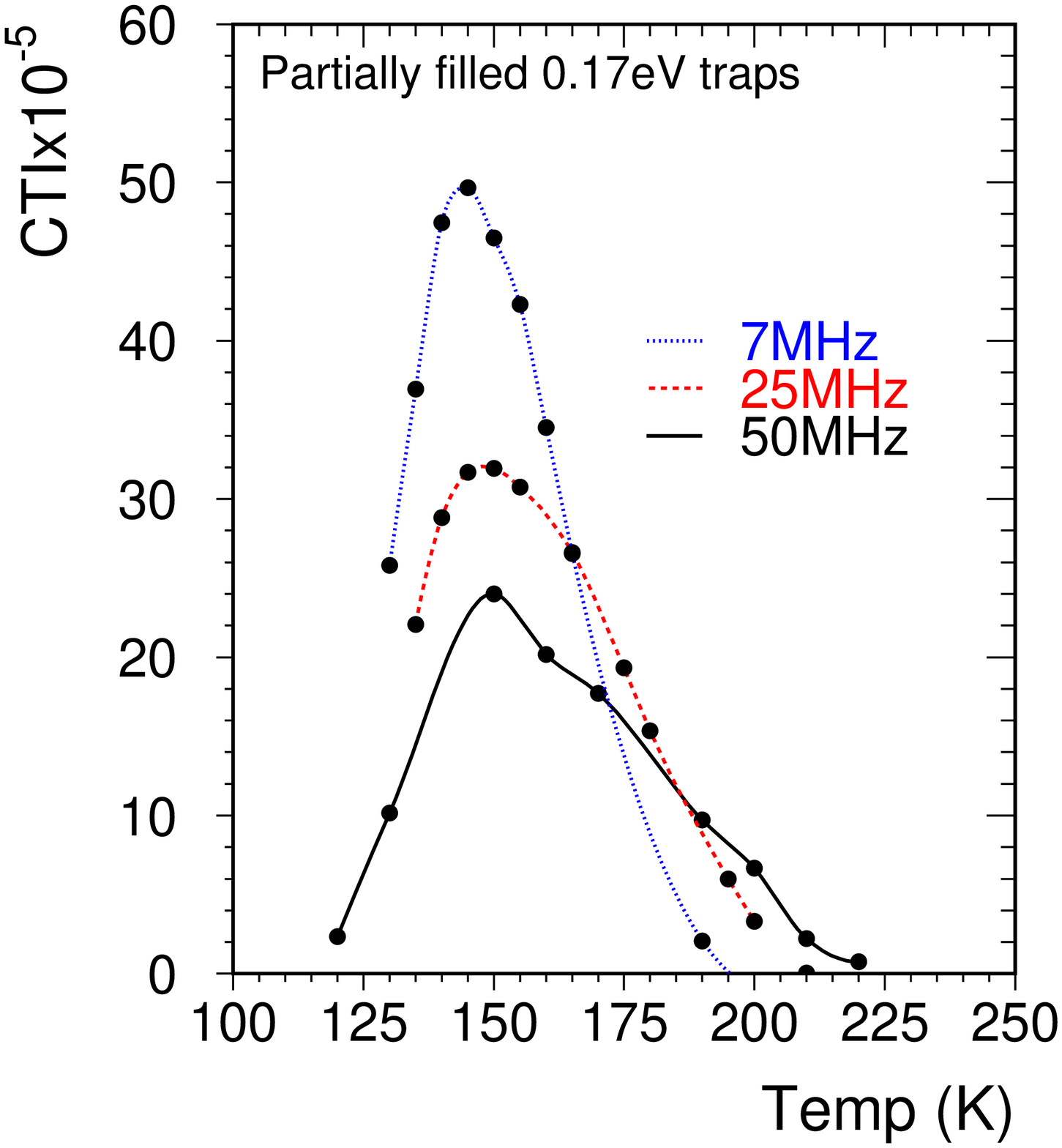}
\end{minipage}
\vspace*{-0.5cm}
\caption{\label{fig:freq}
Left: Frequency dependence for initially empty 0.17 eV traps.
Right: Partially filled 0.17 eV traps.
}
\vspace*{-0.4cm}
\end{figure}

The frequency dependence is shown in Fig.~\ref{fig:freq} for initially empty and partially filled traps. 
For higher readout frequency there is less time to trap the charge, and thus the CTI is reduced near the CTI 
peak region.
At high temperatures, the emission time is so short that trapped charges rejoin the passing signal. 

\section{SIMPLE MODEL}
A simple model of charge trapping was constructed that considered 
the capture and emission of electrons from traps to and from the conduction band.  
These processes are parameterized by two timescales to form a differential rate equation.  
Solution of this equation with relevant boundary conditions allowed this simple model 
to be compared with the full ISE simulation results.

\begin{figure}[htbp]
\begin{center}
\includegraphics[width=8cm,height=4cm]{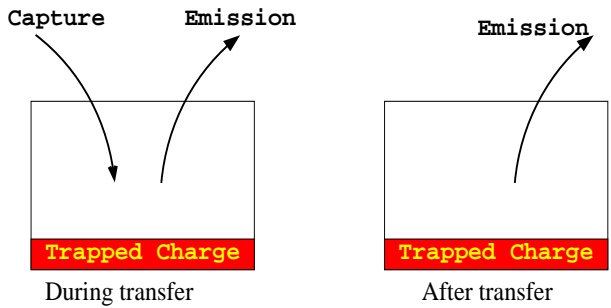} \hfill
\includegraphics[width=8cm,height=6.5cm]{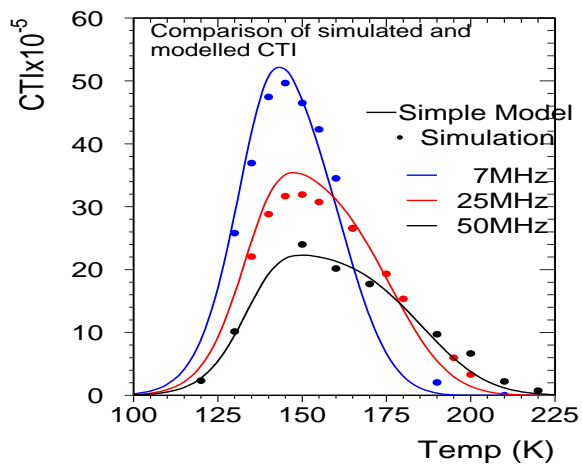}
\vspace*{-0.6cm}
\caption{Left: Traps capture electrons from the signal charge and electrons are emitted from 
the filled traps.
Right: Comparison between simple model (solid lines) and ISE simulation (dots) for three readout 
frequencies.}
\label{fig:compmodel}
\end{center}
\vspace*{-0.3cm}
\end{figure}

Figure~\ref{fig:compmodel} shows the comparison between the full simulation and the simple
model as a function of the temperature. At low temperatures the time for traps to emit 
captured electrons is far longer than the readout time, hence traps remain filled and 
no further electrons can be captured. At high temperatures the emission time is much 
faster than the readout frequency, so captured electrons are released back to the 
conduction band fast enough to rejoin their original signal packet. The CTI value is 
again reduced.
Slower readout frequencies have higher CTI values near their peaks as each pixel has a 
longer occupation time for the signal charge resulting in greater net electron capture.

\section{INITIAL DATA MEASUREMENTS}
\begin{figure}[htbp]
\begin{center}
\begin{minipage}[0.8]{12cm}
\includegraphics[height=6.5cm]{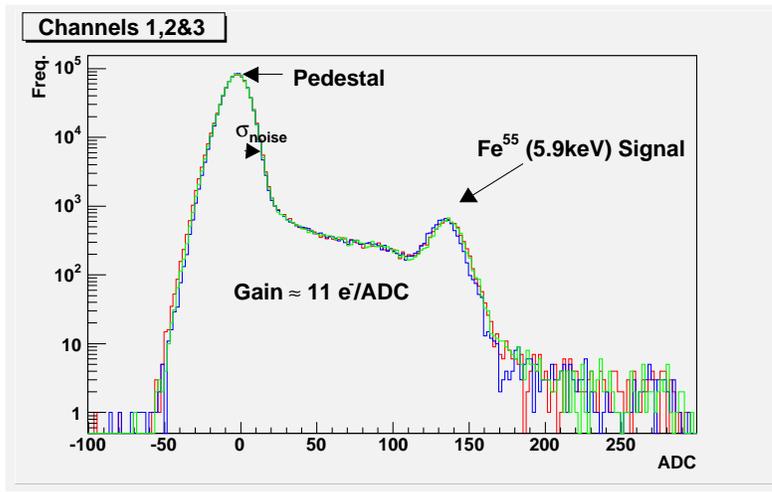}
\end{minipage}
\begin{minipage}[0.2]{4cm}
\caption{Measured signal distribution for an Fe$^{55}$ source.
Noise $\approx60\,{\rm e^-}$,
\vspace*{0.1cm}\\
freq. $=1$\,MHz,
\vspace*{0.1cm}\\
integration time~=~$500$\,ms,
\vspace*{0.1cm}\\
$T\approx-30\,^\circ$C,
\vspace*{0.1cm}\\
2000 frames.
\label{fig:signal}
}
\end{minipage}
\end{center}
\vspace*{-0.6cm}
\end{figure}

For the environment of a future Linear Collider a serial readout for CCDs is no longer an option.  
Instead column parallel technology, using readout electronics for each column, are in 
development to cope with the required readout rate. CPC-1 is a prototype 2-phase CCD 
capable of 25~MHz readout frequency. Initial measurements have been performed on an 
unirradiated device in standalone mode, where four columns of the CCD were connected to
external ADC amplifiers. An Fe$^{55}$ source provides the signal charge (Fig.~\ref{fig:signal}).  
The determination of the CTI involves measuring the charge reduction as a function 
of the pixel number from a known initial charge.
The results so far have shown small CTI values ($<10^{-5}$) for the unirradiated CCD 
under normal operation conditions. It is possible to induce CTI-like effects by reducing the clock 
voltages used to transfer charge. For 1~MHz readout the CTI value 
was observed to increase sharply  below 1.9~V (peak-to-peak).

\subsection{CTI - Event Selection}
\begin{figure}[htbp]
\begin{minipage}[0.45]{8cm}
\includegraphics[height=6.5cm,width=8cm]{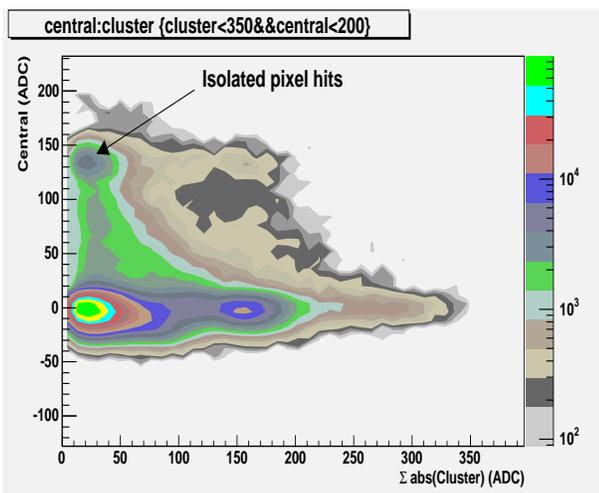}
\end{minipage} \hfill
\begin{minipage}[0.45]{8cm}
\includegraphics[height=6.5cm,width=8cm]{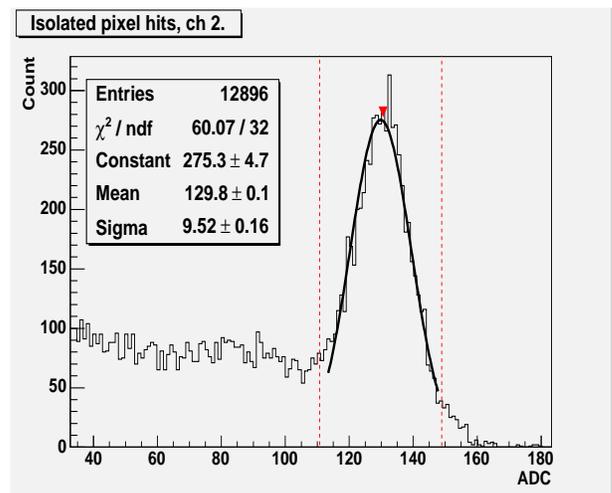}
\end{minipage}
\vspace*{-0.3cm}
\caption{Left: Isolated hit.
Right: Extracted signal peak.
\label{fig:select}
}
\vspace*{-0.3cm}
\end{figure}

The signal is induced by an Fe$^{55}$ source which provides isolated hits of about 1620 electrons 
in order to determine the CTI value. Hits are located using a 3x3 cluster method and selection 
criteria are applied:
Pixel amplitude $> 5\sigma_{\rm noise}$,
$\Sigma_{i=1}^{8} |{\rm cluster}_i|< 8\sigma_{\rm noise}$.
Events are selected within $\pm 2\sigma$ of the signal peak as shown in Fig.~\ref{fig:select}.

\subsection{CTI - Determination}
\begin{figure}[hp]
\vspace*{-0.5cm}
\begin{minipage}[0.45]{8cm}
\begin{center}
\includegraphics[height=6.5cm,width=8cm]{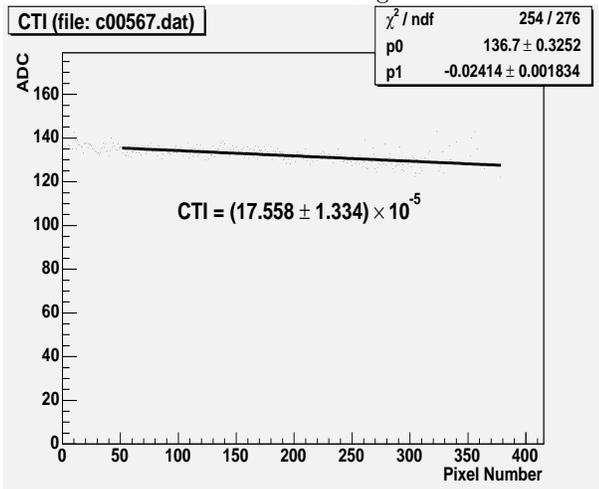}
\end{center}
\end{minipage} \hfill
\begin{minipage}[0.45]{8cm}
\begin{center}
\includegraphics[height=6.5cm,width=8cm]{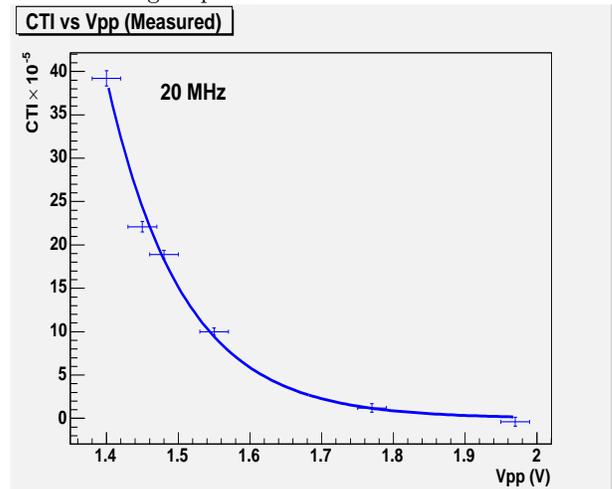}
\end{center}
\end{minipage}
\vspace*{-0.3cm}
\caption{Left: CTI determination.
Right: CTI as function of clock voltage.
\label{fig:clockvoltage}
}
\vspace*{-0.7cm}
\end{figure}

\vspace*{-0.1cm}
The CTI is determined from isolated pixel hits. The distribution of the ADC amplitude $Q$ against the pixel number gives
$
{\rm CTI}=-\frac{1}{Q_0}\frac{{\rm d\,} Q}{{\rm d\,(Pixel)}},
$
where $Q_0$ is the intercept from a straight-line fit. An example for an unirradiated device and low clock 
voltage is given in Fig.~\ref{fig:clockvoltage}.
The decrease of the clock voltage reduces the transfer efficiency which provides a possibility 
to measure CTI values as function of the clock voltage.

\clearpage

\subsection{Dark Current}
\begin{figure}[htbp]
\begin{minipage}[0.45]{8cm}
\includegraphics[height=6.5cm,width=8cm]{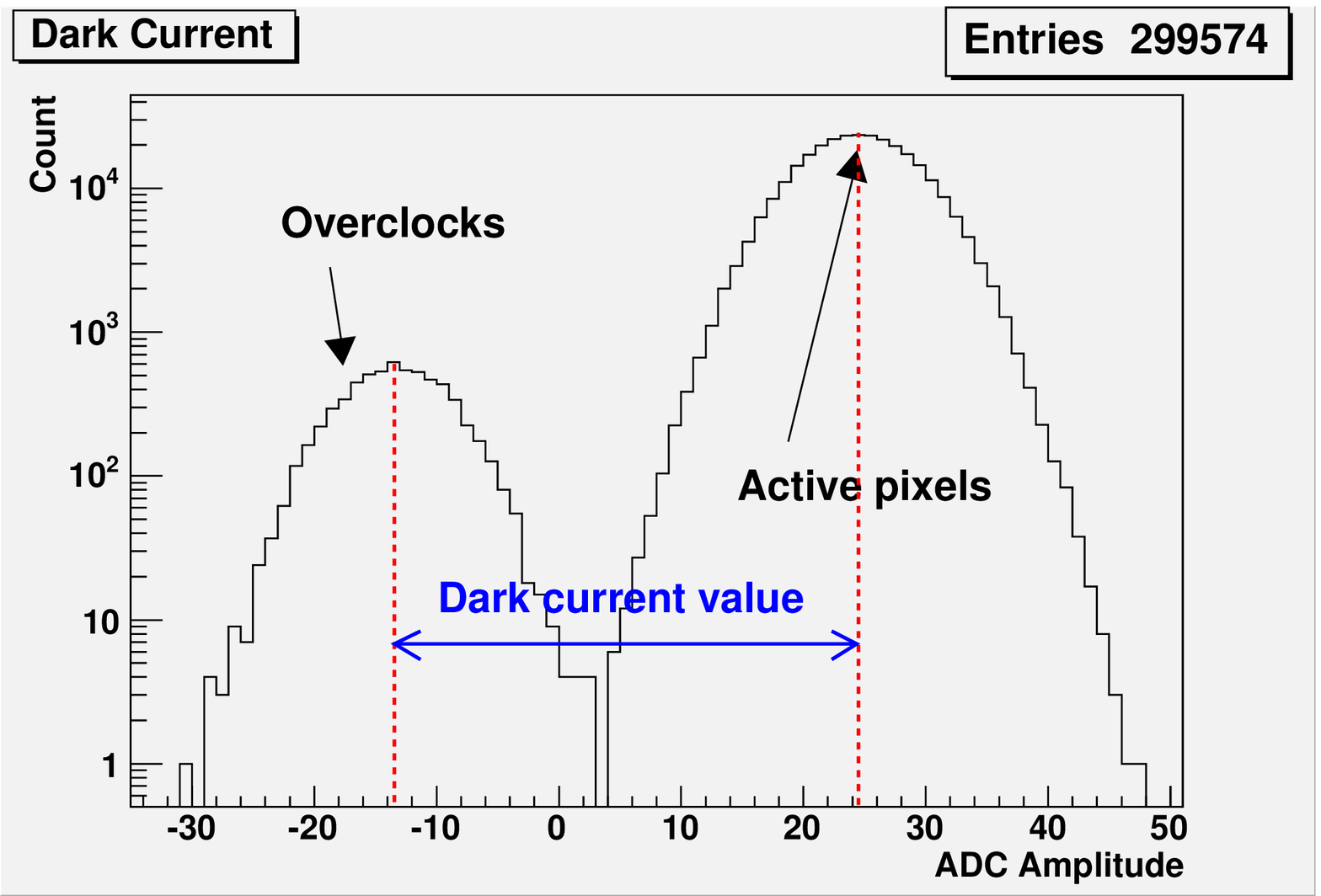}
\end{minipage}\hfill
\begin{minipage}[0.45]{8cm}
\includegraphics[height=6.5cm,width=8cm]{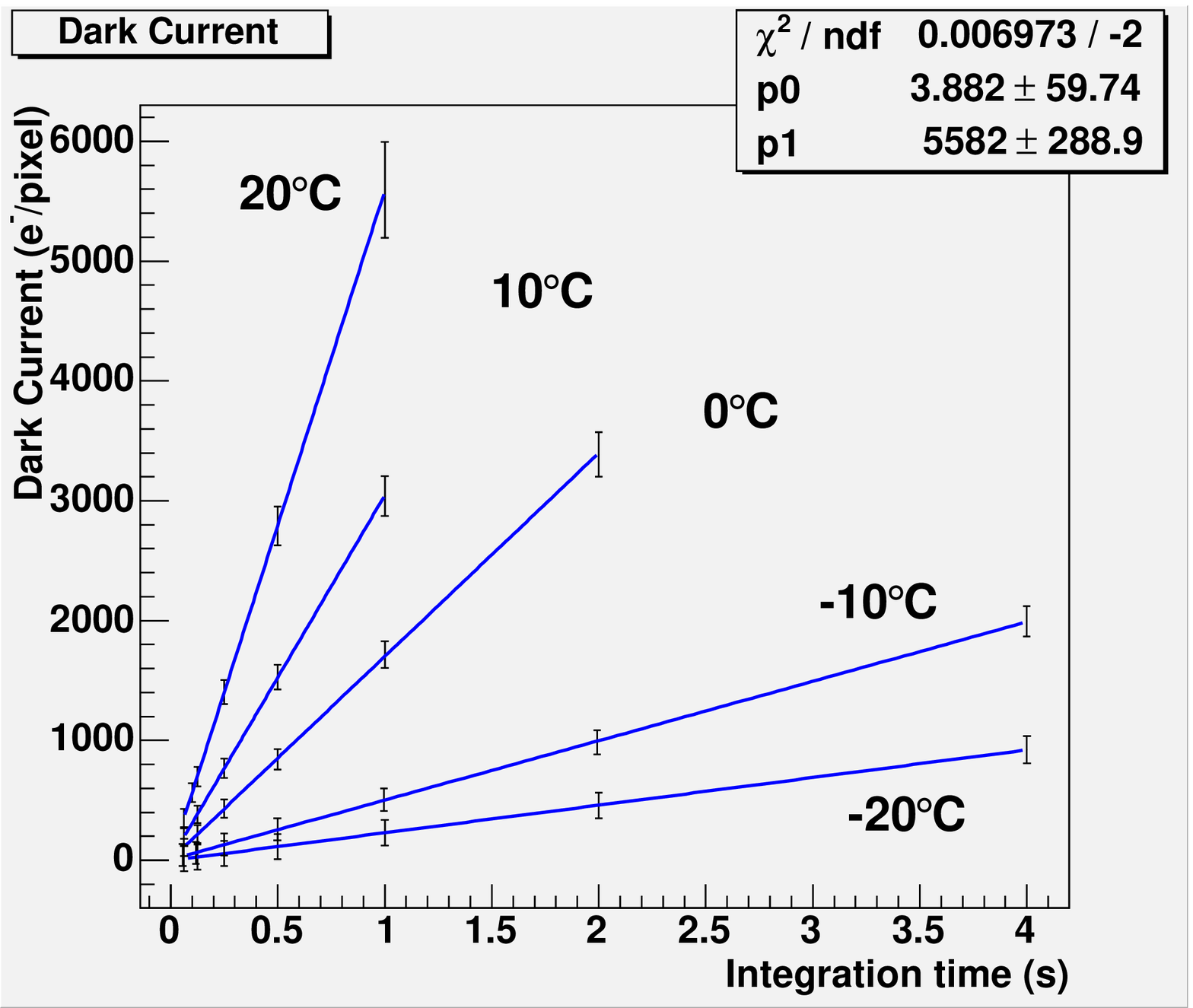}
\end{minipage}
\begin{minipage}[0.45]{8cm}
\includegraphics[height=6.5cm,width=8cm]{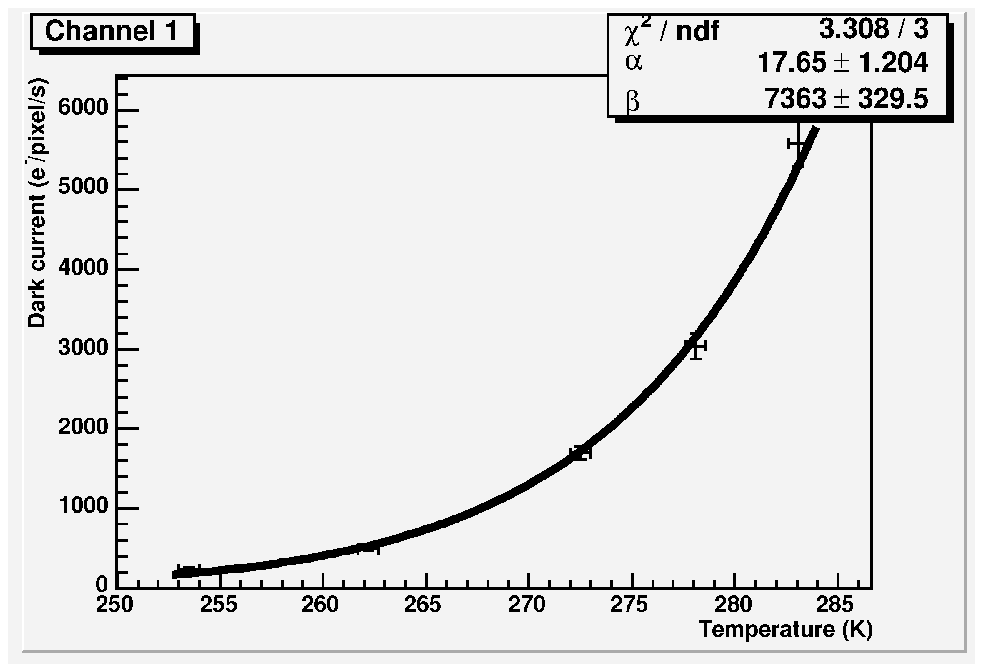}
\end{minipage}\hfill
\begin{minipage}[0.45]{8cm}
\caption{Upper left:  Dark current measurement method.
Upper right: Dark current at different temperatures.
Lower left: Dark current fit to theory expectation.
\label{fig:dc}
}
\end{minipage}
\vspace*{-0.6cm}
\end{figure}

\vspace*{-0.1cm}
Some thermally generated electrons are captured in the potential wells.
The collected charge is proportional to the integration time.
10 overclocks sampled per frame are used as reference level.
The gain (e$^{-}$/ADC) is calibrated from an Fe$^{55}$ source (at each temperature).
From a fit to $J_{\rm dc}=T^3\exp{(\alpha-\beta/T)}$, a uniform dark current characteristics 
is observed across the four channels (Fig.~\ref{fig:dc}).
\vspace*{-0.2cm}

\clearpage

\section{SUMMARY and OUTLOOK}
In summary, radiation hardness effects have been simulated for a serial readout CCD, CCD58.  
A simple model has been developed and compares well with the simulation which  shows the strong 
dependence on the operating conditions of the CCD.
Measurements performed on a prototype column parallel CCD (CPC-1) display as expected no 
measurable CTI for the unirradiated system.
CTI-like effects were induced by reducing the clock voltage to explore the CTI measurement method.
Comparisons with experimental data from irradiated CCDs will be carried out at a test-stand.
First, the expected peak structure will be investigated with experimental data. The test-stand 
at Liverpool University has been shown to operate in the require temperature range. The CCD prototype 
performance will be compared with simulations. Initial experience with this prototype has already 
been gained at the Rutherford Appleton Laboratory (RAL). 
The verification of the simulation results and the tuning of the simulation 
is important for the studies to guide the future CCD development as vertex detector for a future 
Linear Collider.

\begin{acknowledgments}
This work is supported by the Particle Physics and Astronomy Research Council (PPARC) 
and Lancaster University.
I would like to thank Konstantin Stefanov and James Walder for comments on the
manuscript.
\end{acknowledgments}

\end{document}